\documentclass[11pt,showkeys,superscriptaddress]{revtex4-2}
\usepackage{graphics,epsfig}
 
\usepackage{graphicx}
\usepackage{dcolumn}
\usepackage{amsmath}
\usepackage{epstopdf}
\usepackage{hyperref}
\usepackage{color}
\begin{document}
\title{ Exact Solution of Bardeen Black Hole in Einstein-Gauss-Bonnet gravity}
\author{Amit Kumar}
\email{ammiphy007@gmail.com}
\affiliation{Department of Physics, Institute of Applied Sciences and Humanities, GLA University, Mathura 281406, Uttar Pradesh, India}

\author{ Dharm Veer Singh\footnote{Visiting Associate at IUCAA Pune, India}}
\email{veerdsingh@gmail.com}
\affiliation{Department of Physics, Institute of Applied Sciences and Humanities, GLA University, Mathura 281406, Uttar Pradesh, India}
\author{Yerlan Myrzakulov}\email{ymyrzakulov@gmail.com} 
\affiliation{Department of General \& Theoretical Physics, L.N. Gumilyov Eurasian National University, Astana, 010008, Kazakhstan}
\affiliation{ Ratbay Myrzakulov Eurasian International Center for Theoretical Physics, Astana, 010009, Kazakhstan}
 
\author{Gulmira Yergaliyeva}\email{gyergaliyeva1171@gmail.com} 
\affiliation{Department of General \& Theoretical Physics, L.N. Gumilyov Eurasian National University, Astana, 010008, Kazakhstan}
 \author{Sudhaker Upadhyay\footnote{Corresponding author}\footnote{Visiting Associate at IUCAA Pune, India}}
\email{sudhakerupadhyay@gmail.com}
\affiliation{Department of Physics, K.L.S. College, Magadh University, Nawada 805110, India}
 
\affiliation{School of Physics, Damghan University, P.O. Box 3671641167, Damghan, Iran}

\begin{abstract}
We have obtained a new exact regular black hole solution for the EGB gravity coupled with nonlinear electrodynamics in AdS space. The numerical analysis of horizon structure suggests two horizons exist: Cauchy and event. We also study the thermal properties of this black hole, which satisfy the modified first law of thermodynamics. Moreover, we analyse the local and global stability of the black hole. The $P-V$ criticality and phase transition are also discussed. The critical exponents for the present model exactly match the mean field theory. 
\end{abstract} 
\keywords{Bardeen black hole solution; Thermodynamics; Stability; $P-V$ criticality; critical exponents.}
\maketitle
\section{\label{sec:level1}Introduction}
General relativity (GR) is not a complete theory of gravity, as this fails to explain many universe concepts.  For instance, this fails to 
explain the concept of dark matter and dark energy from the first principle. 
In various contexts, the theory of gravitation needs modification, and people have modified it accordingly. Still, there is no complete theory of GR. 
   Lovelock gravity \cite{lav,lav1,lav2} is one of the modified theories of gravity, which mainly emphasises the higher derivative of gravity in higher dimensions.  Recently, the latest developments of modified gravity have been reviewed in cosmology \cite{noj}.
 
 The importance of Lovelock gravity lies in the fact that this is the most general  GR theory, which provides conserved second-order equations of motion in      $D$-dimensions.    Gauss-Bonnet (GB) or    Einstein-Gauss-Bonnet (EGB) gravity is a special kind of Lovelock gravity in higher dimensions, which appears naturally in the low energy effective action of string theory \cite{lan,zwi}.

 Recently, a black hole solution in $4D$ AdS GB massive
gravity is obtained \cite{ds}.
However, a black hole solution in $4D$ AdS  EGB black hole with Yang–Mills field is found in Ref. \cite{ds1}. Here, we are interested in obtaining 
a black hole solution for $5D$ GB massive gravity with nonlinear source term.
 Sakharov and Gliner  \cite{Sakharov:1966, Gliner:1966}  proposed that a de Sitter core with the equation of state $P=-\rho$ or $T_{ab}=\Lambda g_{ab}$ is required to get a non-singular solution, which could provide proper discrimination at the final stage of gravitational collapse. Based on this idea, Bardeen \cite{Bardeen:1968} gave the first regular black hole  and, 30 years later, 
 Ayon-Beato and Gracia find an exact solution coupled to nonlinear electrodynamics \cite{AGB,AGB1,ABG99,AyonBeato:1998ub}. Subsequently, there are many regular black hole solutions obtained. Still, most of this solution are based on Bardeen's proposal \cite{Ansoldi:2008jw,Lemos:2011dq,Zaslavskii:2009kp,Bronnikov:2000vy,Xiang,hc,lbev,Balart:2014cga,singh,fr1,dvs,Singh:2022xgi,Singh:2022dth,Hendi:2015soe,Hendi:2015xya,bks1,bks2,bks3}. The generalization of the regular black hole in EGB gravity \cite{Kumar:2020bqf, Ghosh:2020tgy,Singh:2019wpu,Kumar:2018vsm,Ghosh:2018bxg}, $4D$ EGB gravity \cite{Singh:2020mty,Singh:2020nwo,Singh:2020xju}, massive gravity \cite{Singh:2020rnm} are given.  The rotating counterpart by using the  Newman Janis algorithm \cite{Bambi,Ghosh:2014pba} for  rotating black hole was proposed \cite{Neves:2014aba,Toshmatov:2014nya,Ghosh:2014hea}. Recently, the EGB gravity is simplified in an inflationary theoretical framework, which solves the problem of gravity waves having speed equal to that of light \cite{noj1,noj2}. The compact objects (stars) 
 \cite{noj3} and   primordial gravitational waves  \cite{noj4} are also studied for  GB gravity.

Thermal properties of black holes have been the subject of interest for many years \cite{sud1,sud2,sud3,sud4,sud5}. Black holes are not  only characterised by  
temperature or entropy, but they possess  phase structures and admit critical
phenomena \cite{Kubiznak:2016qmn}. {In fact, phase transition is important in investigating the black hole's properties in extended space. Hawking and Page have studied the first phase transition between  AdS black hole and thermal AdS \cite{h1}. Witten analysed the confinement/deconfinement phase \cite{w1}, and  Chamblin studied the van der Wall phase transition of charged AdS black hole \cite{c1,c2}}. The extended thermodynamics of various black holes are  studied for different types of black holes \cite{Hansen:2016ayo,Hendi:2018sbe,
Hendi:2014kha,Hennigar:2016ekz,rps, Ali:2023ppg,Kumar:2023gjt,Sood:2022fio,Abdusattar:2023xxs,Abdusattar:2023fdm} and the $P-v$ criticality of the FLRW universe are \cite{Abdusattar:2023pck, Abdusattar:2023hlj}. 
Thermodynamics of singular solution for the rotating counterpart of Lee-Wick gravity having a point source in a higher-derivative theory is presented in Ref. \cite{sd1}. Here, we discuss the thermodynamics of $5D$ EGB black hole solution with a nonlinear source. 

The paper is organised as follows. We obtain the  EGB  Bardeen AdS  black hole solution   and also give the relevant equations of EGB gravity coupled to nonlinear electrodynamics with the structure and location of the horizons  in 
 Sec. II. The study of the thermodynamical properties of $5D$ EGB  regular massive black holes is the subject of Sec. III. We end the paper with results and concluding remarks in Sec. V
We use the metric  signature $(-, +,+,+,+)$ with natural units $8\pi G = c = 1$.
\section{Black Holes solutions  In EGB Gravity}\label{sec2}
The  action of EGB gravity  coupled with dual nonlinear electrodynamics in  AdS  spacetime    is written as  
\begin{eqnarray}
S =\frac{1}{2 }\int d^{5}x\sqrt{-g}\left[  {R}
-2\Lambda +\alpha \mathcal{L_{GB}} - {\cal{L}}(F) \right],
\label{action}
\end{eqnarray}
where ${R}$ is the Ricci scalar, $\Lambda$ is the cosmological constant, which can be expressed in terms of the Planck length $l$ as $-6/l^2$, 
$\alpha $  is the GB coupling  and $ \mathcal{L_{GB}}=R_{\mu \nu \gamma \delta }R^{\mu \nu \gamma \delta}-4R_{\mu \nu }R^{\mu \nu }+R^{2}$ is the  Lagrangian density of EGB gravity where $R_{\mu\nu}$ and $R_{\mu\nu\lambda\sigma}$ are  the Ricci and Riemann tensor, respectively. 
Here, Lagrangian ${\cal{L}}$ depends on $F=\frac{1}{4}F_{\mu\nu}F^{\mu\nu}$, where $F_{\mu\nu}=\partial_\mu A_\nu-\partial_\nu A_\mu$ is Maxwell field-strength tensor. $J_\mu$ is the current vector corresponding to the source.   We study the black hole system in terms of the Pleb\'anski tensor $ P_{\mu\nu}$ defined formally as  $ P_{\mu\nu}:=2\frac{\partial {\cal{L}}}{\partial F^{\mu\nu}}=F_{\mu\nu}  {\cal L}_F$, where ${\cal L}_F= \partial {\cal L}/\partial F$.  The nonlinear electrodynamics of the considered system can be obtained alternatively by using the Legendre transformation \cite{AyonBeato:1998ub}: ${\cal{H}}=2F{\cal L}_F-{\cal L}$  which depends on an antisymmetric field $(P=\frac{1}{4}P_{\mu\nu}P^{\mu\nu})$ \cite{a1, AyonBeato:1998ub}. Thus, the Lagrangian for nonlinear electrodynamics can be expressed as 
\begin{equation}
{\cal{L}}=2P {\cal{H}}_P-{\cal{H}} \qquad \text{and}\qquad {\cal{H}}_P=\frac{\partial {\cal H}}{\partial P},
\label{h}
\end{equation}
where  electromagnetic field strength is $F_{\mu\nu}={\cal{H}}_PP_{\mu\nu}$ and ${\cal H}(P)$ is the structure function. 

Extremizing the action (\ref{action}) with respect to metric tensor ($g_{\mu\nu}$) and  potential ($A_{\mu}$) lead to the field equations   \cite{AyonBeato:1998ub, AGB}
\begin{eqnarray}
I_{\mu\nu}\equiv G_{\mu\nu}+H_{\mu\nu}+\Lambda g_{\mu\nu}&=&T_{\mu \nu}=2\left({\cal{H}}_P P_{\mu\lambda}P^{\lambda}_{\nu}-\delta_{\mu\nu} (2P{\cal{H}}_P -{\cal{H}} )\right), \label{eq0} \\
\nabla_{\mu}P^{\mu\nu} &=&\frac{1}{\sqrt{-g}}\partial_{\mu} (\sqrt{-g}P^{\mu\nu} ) = 0.\label{eq}
\end{eqnarray}
Here, only the time component of $J^\nu$ is non-trivial and given by a delta function corresponding to the point source.
 
Here,  the Einstein tensor $G_{ab}$ and the Lanczos tensor $H_{ab}$   are given by \cite{Singh:2019wpu}
\begin{eqnarray}
G_{\mu\nu}&=&R_{\mu\nu}-\frac{1}{2}g_{\mu\nu}R,\\
H_{\mu \nu }& =&-\frac{\alpha }{2}\left[ 8R^{\rho \sigma }R_{\mu \rho \nu
\sigma }-4R_{\mu }^{\rho \sigma \lambda }R_{\nu \rho \sigma \lambda
}-4RR_{\mu \nu }+8R_{\mu \lambda }R_{\nu }^{\lambda }\right.   \notag \\
&& +\left. g_{\mu \nu }\left( R_{\mu \nu \gamma \delta }R^{\mu \nu \gamma
\delta }-4R_{\mu \nu }R^{\mu \nu }+R^{2}\right) \right].
\end{eqnarray}
The explicit form of ${\cal{H}}$ will appear later.
 Here, we want to obtain a $5D$ regular EGB black hole solution in AdS space-time and investigate its properties. For that, we write the static spherically symmetric metric   \cite{Singh:2020rnm}:
\begin{equation}
ds^2=-f(r)dt^2 +\frac{1}{f(r)}dr^2+r^2(d\theta^2+\sin^2\theta d\phi^2+\sin^2\theta\sin^2\phi d\phi^2),
\label{m1}
\end{equation}
where $f(r)$ is an unknown metric function which depends on variable $r$. Therefore, we restrict  the electric field to be
\begin{equation}
F_{\mu\nu}=E(r)(\delta^t_\mu\delta^r_\nu -\delta^t_\nu\delta^r_\mu).
\end{equation}
For the spherically symmetric case, the equation (\ref{eq}) can be 
expressed as
\begin{eqnarray}
\frac{1}{r^2}\frac{\partial}{\partial r}\left(r^2F^{\mu\nu}  {\cal 
L}_F \right)= \frac{\partial}{\partial r}\left( r^2 P^{\mu\nu}
\right)=0. 
\end{eqnarray}
Now,  the term inside the derivative  has to be constant \cite{gra} (we choose an electric charge $e$) 
\begin{equation}
 r^2  P^{\mu\nu}=    -{e}(\delta^t_\mu\delta^r_\nu -\delta^t_\nu\delta^r_\mu) \Longrightarrow P^{\mu\nu}=    -\frac{e}{ r^2}(\delta^t_\mu\delta^r_\nu -\delta^t_\nu\delta^r_\mu),
\label{p}
\end{equation}
and the invariant  $P$ is given by $P =-\frac{e^2}{2r^4}$.
The specific structure-function ${\cal H}(P)$ that depends on the 
invariant $ P$ is given by 
\begin{equation}
{\cal{H}}(P)= \frac{3}{2 s e^4}\left(\frac{ \sqrt{-2 e^2 P} }{ 1+
\sqrt{-2e^2P} }\right)^{7/3},
\label{matter}
\end{equation}
where $s$ is the   free parameter related to the  ADM  mass $M$ and 
charge  $e$  by $s =|e|/2M$.  The structure-function (\ref{matter}) 
justifies  the reasonable conditions needed
for   nonlinear electrodynamics, as this goes over to the Maxwell 
linear electrodynamics, $ {\cal{H}}(P) \to P$ for the weak fields 
($P<<1$) and also satisfying the weak energy condition, which 
requires ${\cal H}<0$ and $ {\cal{H}}_P>0$ \cite{AGB1,lbev,
Balart:2014cga}.

 With the help of equations (\ref{eq0}) and (\ref{p}), we derive the non-zero components of the energy-momentum tensor as
\begin{eqnarray}
&&T^t_t=T^r_r= \frac{3 e^3 M}{(r^3+e^3)^{7/3}} ,\nonumber\\
&&T^{\theta}_{\theta} =T^{\phi}_{\phi} =T^{\psi}_{\psi} =\frac{2 e^3 M (3e^3-4r^3)}{(r^3+e^3)^{10/3}}.
\end{eqnarray}
  Using the metric ansatz (\ref{m1}), the non-vanishing components of the Einstein field equation become
\begin{eqnarray}
&&I_t^t =I^r_r=(4\alpha f'(r)-2 r)(f(r)-1)-r^3f'(r)=\frac{6M e^3 }
{(r^3+e^3)^{7/3}},\nonumber \\
&&I^{\theta}_{\theta} =I^{\phi}_{\phi} =I^{\psi}_{\psi}  =\frac{2rf'+f+
\frac{r^2}{2}f''+2\alpha(f''+f'^2+f'')-1}{r^2}=\frac{2 e^3 M (3e^3-4r^3)}
{(r^3+e^3)^{10/3}},
\label{eom1}
\end{eqnarray}  
where  single   and double prime are the first and second derivatives, with  
respect to $r$, respectively. 
The above equations lead to the  metric function
\begin{equation}
f(r)=1+\frac{r^2}{4\alpha} \left(1\pm \sqrt{1+\frac{8\alpha M}{(r^3 + 
e^3)^{4/3}}-\frac{8\alpha}{l^2}}\right).
\label{sol1}
\end{equation}
 Here, one can see that the solution is characterised by the parameter mass, 
charge and GB coupling constant. {Here, we note that there are   two branches (  $+ ve$ and $- ve$ branch) of solution (\ref{sol1}).  
In the $M=0$ limit, the black hole solution (\ref{sol1}) becomes
\begin{equation}
f(r)=1+\frac{r^2}{4\alpha} \left(1\pm \sqrt{1-\frac{8\alpha}{l^2}}\right).
\end{equation}
When $\alpha > 0$  then $8\alpha/l^2\leq 1$, there is no black hole solution beyond this limit.  Thus, the action \ref{action} has two solutions with effective cosmological constants $l^2_{eff}=\frac{l^2}{4} \left(1\pm \sqrt{1-\frac{8\alpha}{l^2}}\right)$ . However, for $8\alpha/l^2 = 1$, both the solutions coincide, and therefore, the theory has a unique AdS vacuum.

{ 
The studies of the obtained black hole solutions will help to understand the difference between   the other black hole solutions \cite{Singh:2020mty,bks2} of the similar class of EGB gravity. 

\begin{enumerate}
\item The source of the obtained black hole solution is unique as 
this is different from the previous solutions \cite{Singh:2020mty,bks2}. The departure of Lagrangian density from the other  
sources of black hole solutions is more evident from the order expansion:
\begin{equation}
    {\cal L}(P)=\frac{3}{2se^4}  (-2e^2P)^{7/3}\left(1-\frac{7}{3} \sqrt{-2e^2P} +\frac{70}{9}(\sqrt{-2e^2P})^2  +O[P^2]\right).
\end{equation}

\item The black hole solution is the extension of the Bardeen black hole. In the weak field (large $r$) limit, the black hole does not reduce to Reissner-Nordstrom black hole, whereas  the non-Bardeen class black hole smoothly goes over to it. For small $r$,  the black hole solution reduces to
\begin{equation}
    f(r)= 1+\frac{r^2}{4\alpha}\left[1\pm \sqrt{1+\frac{8M\alpha}{r^4}+\frac{32 M \alpha e^3}{r^7}-\frac{8\alpha}{r^2}+ O[e^6]}\right].
\end{equation}
\noindent The  black hole solution  (\ref{sol1}) reduces to the Boulware-Deser  black hole \cite{7}, $5D$ Bardeen black hole  \cite{Ali:2018boy} and 
Schwarzschild-Tangherilin black hole in the limit of $e=0$,   $\alpha\to 0$ and  $e=0,\alpha \to 0$, respectively. 
\end{enumerate} 
}

When $\alpha < 0$ , the solution (\ref{sol1}) still becomes  AdS if one takes the $- ve$ signature and $dS$ if one takes the $+ ve$ signature.  From the vacuum case, the solution (\ref{sol1}) with both signs seems reasonable, from which we cannot determine which
sign in (\ref{sol1}) should be adopted. Then, Boulware and Deser showed that the solution with $+ ve$  branch is unstable and the solution is asymptotically an  AdS  Schwarzschild solution with negative gravitational mass, indicating the instability. The solution (\ref{sol1}) with $- ve$ branch is stable and the solution is asymptotically a Schwarzschild solution. This indicates that the  $+ ve$  branch is of less physical interest. Henceforth, we adopt the negative branch of the solution for further analysis. }
  
This solution  behaves at large $(r\to
\infty)$ and small $(r\to 0)$ distance as following:
\begin{eqnarray}
&&f(r)=1-\frac{M}{r^2},\qquad\qquad r \to\infty\nonumber\\
&&f(r)=1-{\Lambda_{eff}\,r^2}, \qquad \,\,r \to0
\end{eqnarray}   
where  $\Lambda_{eff}=\left(1-\sqrt{1+{8M\alpha}/{e^4}-8\alpha/l^2}\right)/
{4\alpha}$. This implies that the EGB Bardeen  AdS  black hole solution has a central 
de Sitter core.

 The numerical analysis of horizon condition imparts that there exists non-zero    $\alpha$ and $e$ for which metric function $f(r)$ is minimum. However, the horizon condition of metric admits two possible  roots for horizon radius  $r_-$ and $r_+$ that correspond  to the Cauchy and event horizon, respectively.

\begin{figure*}[ht]
\begin{tabular}{c c c c}
\includegraphics[width=.525\linewidth]{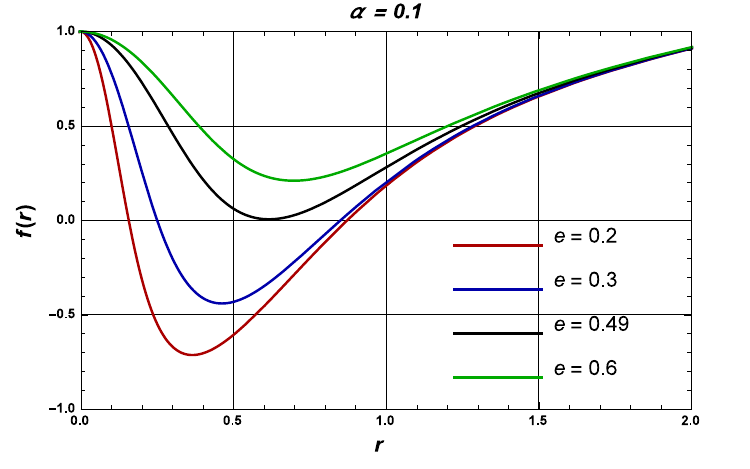}
\includegraphics[width=.525\linewidth]{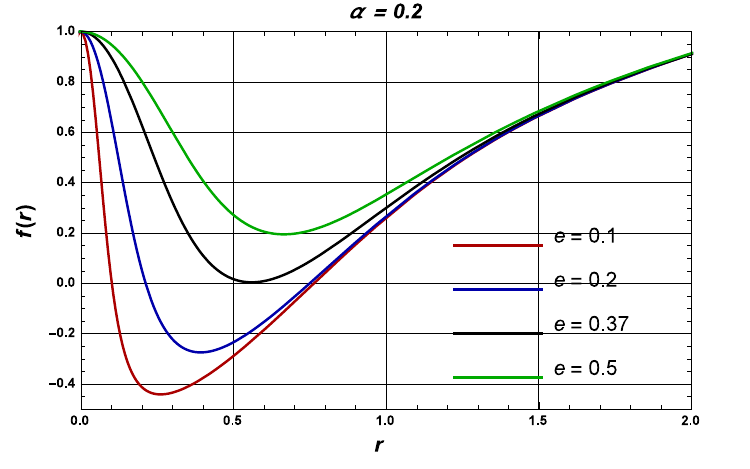}
\end{tabular}
\caption{  $f(r)$  vs $r$ corresponding to GB parameter $\alpha=0.1$ (left)
and $\alpha=0.2$  (right) for different values of charge  $e$.  }
\label{figh1}
\end{figure*}

We tabulated the numerical values  of  the inner and outer horizons corresponding 
to various parameter in Tab. \ref{tabh1}.  
\begin{center}
\begin{table}[h]
\begin{center}
\begin{tabular}{| l| l| r |l| r| l| r |l | r  }
\hline 
\multicolumn{1}{|c }{ }&\multicolumn{1}{c}{ $\alpha=0.1$}&\multicolumn{1}{c}{}&\multicolumn{1}{c|}{  }&\multicolumn{1}{c}{ }&\multicolumn{1}{c}{ $\alpha=0.2$  }&\multicolumn{1}{c}{}{ }&\multicolumn{1}{c|}{} \\
\hline
\hline
\multicolumn{1}{|c|}{ $e$} & \multicolumn{1}{c|}{ $r_-$ } & \multicolumn{1}{c|}{ $r_+$ }& \multicolumn{1}{c|}{$\delta$}&\multicolumn{1}{c|}{$e$}& \multicolumn{1}{c|}{$r_-$} &\multicolumn{1}{c|}{$r_+$} &\multicolumn{1}{c|}{$\delta$}   \\
\hline

\,\,\,\, $ 0.493$ $(e_c)$\,\, &  \,\,0.624\,\,  &\,\,0.624\,\,&\,\,0\,\,&\,  $0.373$ $(e_c)$ \, &\,\,0.566\,\,&\,\,0.566\,\,&\,\,0\,\,
 \\
\
\,\, $0.40$\,\,& \,\,0.37\,\, &\,\,  0.81\,\,& \,\,0.44\,\,&\,$0.25$\,\, \ \ \ &\,\,0.27\,\,&\,\,0.74\,\,&\,\,0.47\,\,
\\
\
\,\, $0.45$\,\, & \,\,0.45\,\, &\,\,  0.76\,\,& \,\,0.31\,\,&\,$0.25$\,\, \ \ \ &\,\,0.35\,\,&\,\,0.71\,\,&\,\,0.36\,\,
\\
\hline
\end{tabular}
\end{center}
\caption{Radius of inner and outer horizons and $\delta=r_+-r_-$ for different values of charge $e$.}
\label{tabh1}
\end{table}
\end{center}
 The metric function with respect to horizon radius is depicted in Fig. \ref{figh1}.
From   Fig. \ref{figh1} and TAB. \ref{tabh1},  one can see that for a critical value of charge (say $e_c$) the minimum of metric function is zero
(black line curve) and  there exists a   degenerate horizon. However, for $e < e_c$  the black hole have two horizons $r_{\pm}$ correspond to the non-extremal black hole and  for $e>e_c$ there is no horizon, i.e., no black holes. It is noticed that the  size of the black hole decreases with the  increase in charge and decrease in the GB coupling. This means that the effect of charge and GB coupling are opposite to each other. We can  also see that the radius of the event horizon  increases with the $e$ and decreases with the $\alpha$.

The obtained black hole solution is singular if curvature invariants (Ricci scalar $R$, Ricci square $R_{\mu\nu}R^{\mu\nu}$ and Kretshmann scalar $R_{\mu\nu\lambda\sigma}R^{\mu\nu\lambda\sigma}$)   diverge and is regular if   curvature invariants converge. Here, for this system, the curvature invariants are calculated by
\begin{eqnarray}
 \lim_{r\to 0} R &=&-\frac{5}{\alpha}+\frac{5}{\alpha}\sqrt{1+\frac{8M\alpha}{e^4}-\frac{8\alpha}{l^2}},\nonumber\\
 \lim_{r\to 0}R_{\mu\nu}R^{\mu\nu}&=&\frac{10}{\alpha^2}+\frac{40}{e^4 \alpha }-\frac{10}{\alpha^2}\sqrt{1+\frac{8M\alpha}{e^4}-\frac{8\alpha}{l^2}},\nonumber\\
 \lim_{r\to 0} R_{\mu\nu\lambda\sigma}R^{\mu\nu\lambda\sigma}&=&\frac{5}{\alpha^2}+\frac{20}{e^4 \alpha}-\frac{5}{\alpha^2}\sqrt{1+\frac{8M\alpha}{e^4}-\frac{8\alpha}{l^2}}.
\label{inv}
\end{eqnarray}
Here, we find that for $M \neq 0 $, the invariants are well-behaved everywhere including the origin. Thus, the EGB Bardeen  AdS  black hole is a regular (non-singular) black hole.

 The obtained black hole solution is  regular black holes with a de Sitter core   at the Cauchy horizon, $r_0=r_-$, which is a null surface.  The interior region, $r < r_0 = r_-$, is uncharged, satisfying a de Sitter equation of state, where $r_0$ denotes the surface boundary.  

Let us now discuss the features of the three different
regions of this regular black hole:

(I)   In the region,  $r<r_0=r_-= R$, the
region is uncharged.

(II) In the region, $r=r_0=r_-=R$, one
has an electrically charged but massless layer.  

(iii) The   region, $r>r_0=r_-= R$, is the Reissner-Nordstr\"om like vacuum.
 \section{Brief comparison with other similar solutions}
 Recently, in Ref. \cite{shu}, a black hole solution for $4D$ EGB gravity with different non-linear source is discussed and  it is shown there that in the large asymptotic limit, the negative branch of the solution which corresponds to Schwarzschild AdS black hole, whereas the positive branch  is not physical as it corresponds to the negative mass. However, in vanishing GB parameter, the solution corresponds  to Hayward Ads black holes.
 
 In another work \cite{shu1}, a regularized 4D EGB gravity coupled to the different nonlinear electrodynamics  is studied and this   black hole undergoes a phase transition twice. However,   the Hayward   and Born-Infeld   EGB black holes undergo phase transition only once.
\section{Thermodynamics}\label{sec3}
In this section,  we derive  thermodynamical quantities of EGB Bardeen  AdS  black holes solution satisfying the first law of thermodynamics. 
The mass of the black hole on the horizon is calculated by 
\begin{eqnarray}
 M_+=(e^3+r_+^3)^{\frac{4}{3}}\left(\frac{ r_+^2+2\alpha }{r_+^4}+\frac{1}{l^2}\right).\label{mass}
 \end{eqnarray}
Using the standard definition, the Hawking temperature corresponding to the given black hole solution is
 given by
\begin{eqnarray}
 T_+=\left.\frac{f'(r)}{4\pi}\right|_{r=r_+}=\frac{1}{4\pi r_+(e^3+r_+^3)(r_+^2+4\alpha)}\left[\frac{2r_+^7}{l^2}+ {2r_+^5-2e^3(r_+^2+4\alpha)} \right].\label{temp}
\end{eqnarray}
From this expression, it is evident that the  temperature of the EGB Bardeen  $AdS$ black hole is more general than  that of EGB black hole \cite{7},  Bardeen black hole \cite{Ali:2018boy}  and Schwarzschild-Tangherilini black hole  as the corresponding values of temperature can be achieved by just taking the limiting value of $e=0$,  $\alpha = 0$  and    $\alpha = e= 0$, respectively.

To study the nature of temperature corresponding to different parameters, we plot  temperature concerning $r_+$ for different $e$ and $\alpha$ in Fig. \ref{figt2}.  
From the diagram, we see  that the temperature takes  a maximum value ($T^{Max}_+$) corresponding to the minimum metric function. 
{ \begin{figure*}[ht]
\begin{tabular}{c c c c}
\includegraphics[width=.525\linewidth]{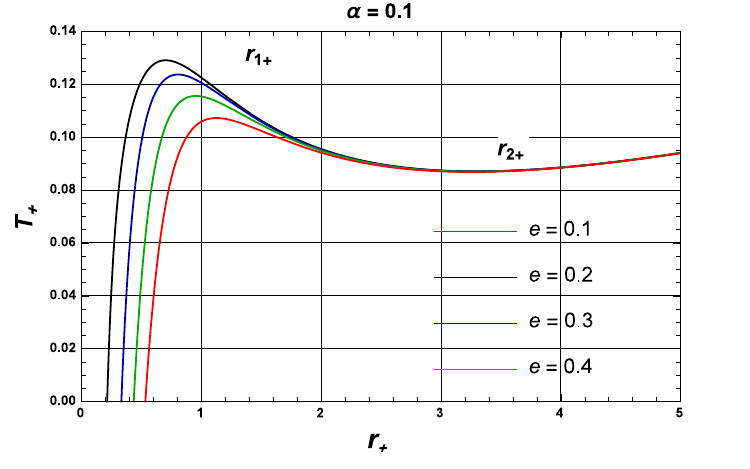}
\includegraphics[width=.525\linewidth]{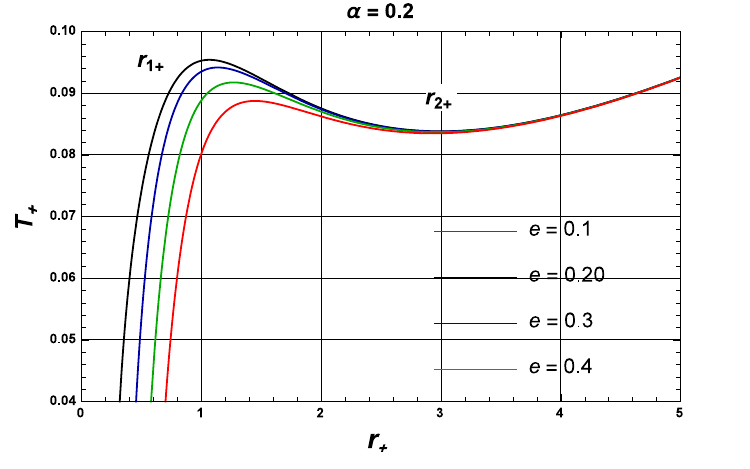}
\end{tabular}
\caption{The plot of temperature Vs. $r_+$ with different value of electric charge for $\alpha=0.1$ and $\alpha=0.2$ with fixed value of $e$. }
\label{figt2}
\end{figure*}
The numerical values of maximum temperature ($T^{Max}_+$) corresponding to  $e$ and $\alpha$ are tabulated in Tab. \ref{tab:temp}.  It is obvious from this table that a   Hawking temperature takes maximum values  at the critical horizon radius. In the extremum limit, (where the Cauchy and event horizons  coincide) the temperature vanishes.
\begin{center}
\begin{table}[h]
\begin{center}
\begin{tabular}{|l|l| r| l| r| l| r| l| r| r| r|}
\hline
\multicolumn{1}{|c|}{ }&\multicolumn{1}{c}{ }&\multicolumn{1}{c}{ }&\multicolumn{1}{c }{$\alpha=0.1$  }&\multicolumn{1}{c}{  }&\multicolumn{1}{c|}{ }&\multicolumn{1}{c}{  }&\multicolumn{1}{c}{ $\alpha=0.2$}&\multicolumn{1}{c }{}&\multicolumn{1}{c }{} &\multicolumn{1}{c| }{}\\
\hline
\,\,$e$\,\, &\,\,0 \,\,&  \,\, 0.40\,\, &\,\,0.45\,\, &  \,\,0.493\,\, &\,\,0.55\,\,&\,\,0\,\,&\,\,0.25\,\,&\,\,0.30\,\,&\,\,0.373\,\,&\,\,0.45\,\,
\\ \hline 
\,\,$r_c $\,\, &\,\,0.63\,\,&  \,\, 0.104\,\, &\,\,0.108\,\, &  \,\,1.11\,\, &\,\,1.19\,\,&\,\,0.897\,\,&\,\,0.996\,\,&\,\,1.088\,\,&\,\,1.36\,\,&\,\,1.39\,\,
\\ \hline
\,\,$T_+^{Max}$\,\,&\,\,0.092\,\,&\,\,  0.097\,\, & \,\,0.090\,\,& \,\,  0.087\,\, &\,\,  0.082\,\,&\,\,0.088\,\,& \,\,0.095\,\,&\,\, 0.0812\,\,&\,\,0.0786\,\,&\,\,0.0743\,\,
\\
\hline
\end{tabular}
\end{center}
\caption{The maximum Hawking temperature ($T_+^{Max}$) at critical radius ($r_c $) for different values of  charge ($e$).}
\label{tab:temp}
\end{table}
\end{center}
}

The first law of black hole thermodynamics follows 
\begin{equation}
dM_+ = T_+ dS_+ + \Phi de,
\end{equation}
where $\Phi$ is the electric potential corresponding to the electric charge $e$. 
Now, we calculate the entropy for the obtained  black hole solution (\ref{sol1})  
\begin{equation}
S_{+} =\int T^{-1}_+ d M_+ = \int \frac{1}{T_+}\frac{\partial M_+}{\partial r_+}d r_+ = \frac{4\pi r^{3}_+}{3}\left[1+\frac{12\alpha}{r^{2}_+}-\frac{e^4}{r^{6}_+}(3r^{2}_+ + 4\alpha)\right].\label{18}
\end{equation} 
Here, we note that the   two additional terms in    Eq. (\ref{18}) extend  the entropy and the usual area-law $S=\frac{A}{4}$ is no longer valid. In the absence of charge $e$, the above entropy    (\ref{18}) matches exactly to entropy of EGB black hole \cite{7}.

It is well-known  that  entropy of the  regular black hole does not obey the area-law \cite{Ansoldi:2008jw,wald93,Bronnikov:2000vy,singh,fr1}
because the energy-momentum tensor already includes the mass of the black hole. To remove the discrepancy, Ma and Zhao proposed the corrected form of the first-law  of black hole thermodynamics for regular black holes \cite{ma14} by introducing an extra factor. The modified first-law thermodynamics  \cite{ma14,Singh:2022xgi,M2}  reads
\begin{equation}
\xi(M_+,r_+)dM=T_+\,dS +  \Phi \,de,
\label{mod}
\end{equation}
 where  $\xi(M_+,r_+)$ is given by
 \begin{equation}
\xi(M_+,r_+)=1+4\pi \int_{r_+}^{\infty}r_+^2\frac{\partial T^t_t}{\partial M} dr_+=\frac{r_+^4}{(r_+^3+e^3)^{4/3}}.
\end{equation}
One can easily check that  with this value of $\xi(M_+,r_+)$  the entropy of the modified first-law of thermodynamics  follows the area-law.

\section{Local  and Global Stability}\label{sec4}
The local and global stability of the thermodynamical system can be emphasized by studying the heat capacity ($C_+$) and Gibbs free energy ($G_+$) of the system. For instance, the thermodynamic system remains stable when $C_+ >0 $ or $G_+<0$ and unstable when $C_+<0$ or $G_+>0$. The  heat capacity of the thermal system is defined by  
\begin{equation}
C_+ =\frac{\partial M_+}{\partial T_+} = \left(\frac{\partial M_+}{\partial r_+}\right)\left(\frac{\partial r_+}{\partial T_+}\right).
\label{heat1}
\end{equation}
Substituting the values of mass (\ref{mass}) and temperature
(\ref{temp})   into (\ref{heat1}), the heat capacity for the  the Bardeen black hole in EGB gravity has the following expression:
\begin{equation}
C_+ =\frac{4\pi \beta^2\left[r_+^5(l^2+2r_+^2)-e^2l^2\beta\right]}{r_+^3\left[e^6l^2\beta^2+r_+^8(-l^2\beta_1^2+2r_+^2\beta_2)+e^3(8r_+^7\beta_3+l^2(6r_+^5 \beta_4+64r_+^3\alpha^2))\right]},
\label{hc1}
\end{equation}
where $\beta=r_+^2+4\alpha$, $\beta_1=r_+^2-4\alpha$,  $\beta_2=r_+^2+6\alpha$  and $\beta_2=r_+^2+8\alpha$.
\begin{figure*}[ht]
\begin{tabular}{c c c c}
\includegraphics[width=.525\linewidth]{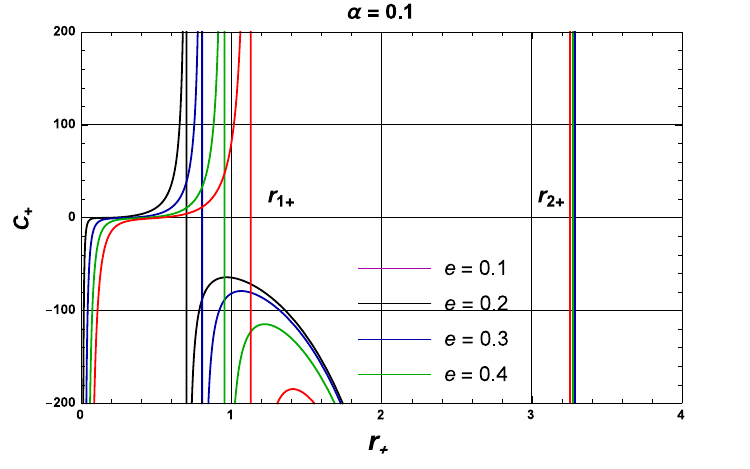}
\includegraphics [width=.525\linewidth]{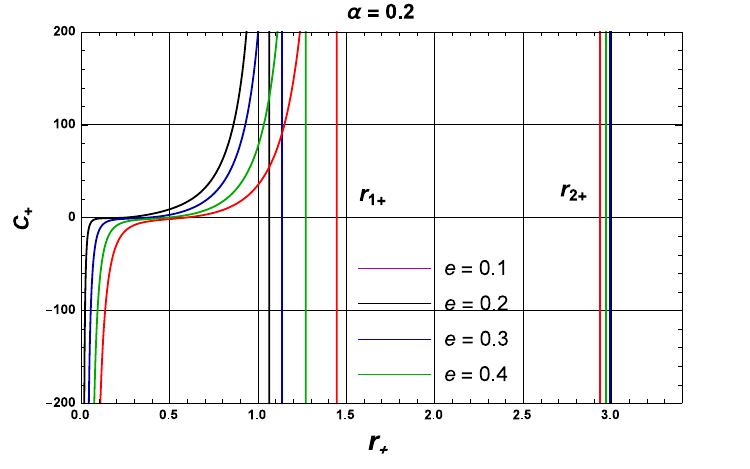}
\end{tabular}
\caption{The plot of  heat capacity $C_+$ Vs. $r_+$ with different value of electric charge for $\alpha=0.1$ and $\alpha=0.2$ with fixed value of $e$. }
\label{fig4}
\end{figure*}

\noindent The  heat capacity with respect to $r_+$ are plotted  in   Fig. \ref{fig4} for different values of $e$ and $\alpha$ which suggests that the heat capacity is divergent at the $r_+=1$ and $r_+=3.2$ for $\alpha=0.1$, however, at $r_+=1.5$ and $r_+=3.0$ for $\alpha=0.2$ where corresponding  temperature has local maxima and local minima. From the Fig. \ref{fig4},  we can see clearly that the black hole experiences phase transition twice, firstly, for smaller stable black holes to larger unstable black holes and, secondly, for smaller unstable black holes to larger stable black holes. The radii $r_+$ increases with both $\alpha$ and $e$.

Now,   to study the  Gibbs free energy, we first estimate using the following expression:
\begin{equation}
G_+ = M_+ - T_+ S_+.
\end{equation}
Substitute the values  of $M_+$, $T_+$ and $S_+$, we obtain 
\begin{equation}
G_+ = (e^3+r_+^3)^{4/3}\left(\frac{ r_+^2+2\alpha }{r_+^4}+\frac{1}{l^2}\right)-\frac{2r_+^2}{3(e^3+r_+^3)(r_+^2+4\alpha)}\left(\frac{r_+^7}{l^2}-{(r_+^5-e^3(r_+^2+4\alpha))}\right).\label{gg}
\end{equation}
At the critical temperature, the Gibbs free energy   vanishes. Thus, the critical temperature  can be calculated   by using $G_+=0$ as
\begin{equation}
T_{min} = \frac{3}{4\pi} \frac{(e^3+r_+^3)^{4/3}}{r_+^3}\left(\frac{(r_+^2+2\alpha)}{r_+^4}+\frac{1}{l^2}\right).
\end{equation}

The black hole is said to be globally stable when $T_+ > T_C$. However, $T_+ < T_C$ describes the global instability of the black hole.  
\begin{figure*}[ht]
\begin{tabular}{c c c c}
\includegraphics[width=.525\linewidth]{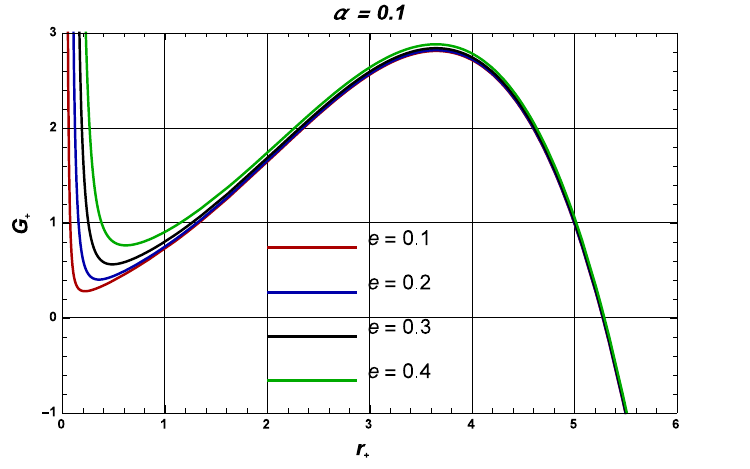}
\includegraphics[width=.525\linewidth]{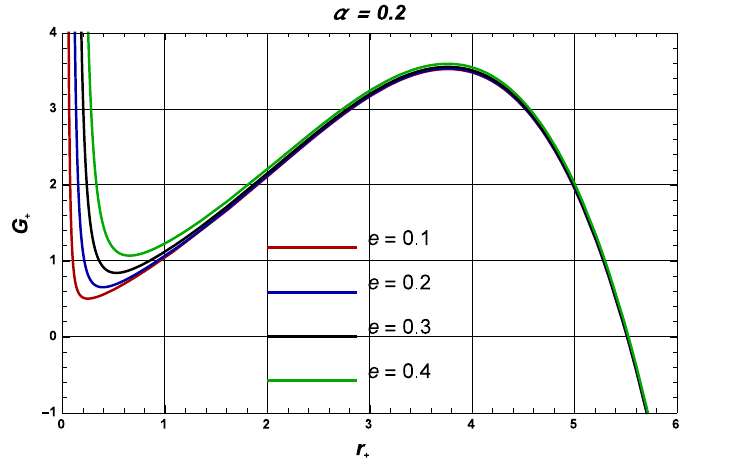}
\end{tabular}
\caption{The plot of  Gibbs free energy $G_+$ Vs. $r_+$ with different value of electric charge for $\alpha=0.1$ and $\alpha=0.2$ with fixed value of $e$. }
\label{fig5}
\end{figure*}

{
Generally, the Gibbs free energy demonstrates that black holes with  negative values of Gibbs free energy are more thermodynamically stable and unstable for  its positive counter parts. 
To study the nature of Gibbs free energy, we plot  (\ref{gg}) in Fig. \ref{fig5} for different values of $e$ and $\alpha$. From the figure, we see that the free energy has local minima and local maxima at horizon radii $r_+=0.1$ and $r_+=3.8$, respectively, where the heat capacity diverges (see Fig. \ref{fig4}) and the temperature attains the extreme values (see Fig. \ref{figt2}).  The Hawking-Page first order
phase transition occurs at $r_+ = r_{HP}$, where the free energy turns negative viz., $r_{HP} > rb
+ $. Thus the larger black holes,
with horizon radii $r_+ = r_{HP}$, are thermodynamically globally stable. However, at very small horizon radii, the Hawking
temperature is negative and hence not physical for global stability point of view. This is exactly in accordance with the Hawking-Page phase transition in general relativity. We find that the black hole solution is favoured globally with respect to the thermal
AdS background solution as $G_+ < 0$ for large $r$.

}
\section{Phase Transition}\label{sec5} 
 \noindent In this section,  the  $ P-v$ criticality and phase transition for  the Bardeen  AdS black hole  EGB gravity.  The cosmological constant is related to the thermodynamic pressure as $\Lambda=-8\pi P$. Thus, the equation of state (EoS) for pressure  can be obtained by using the Eq. (\ref{temp}) as
\begin{eqnarray}
P_+=\frac{3}{8\pi r_+^7}\left(e^3(r_+^2+4\alpha)-r_+^5\right)+\frac{3T}{4r_+^6}\left((r_+^3+e^3)(r_+^2+4\alpha)\right).
\label{pv4}
\end{eqnarray}
  The mass of  the black hole has the interpretation of the enthalpy of the thermodynamic system. Thus, the thermodynamic volume is calculated by
\begin{eqnarray}
 V=\left(\frac{\partial M_+}{\partial P_+}\right)_{S_+}= \frac{4\pi}{3}{\left(r_+^{3}+e^{3}\right)^{\frac{4}{3}}}.
\label{pv3}
\end{eqnarray}

\begin{figure*}[ht]
\begin{tabular}{c c c c}
\includegraphics[width=.525\linewidth]{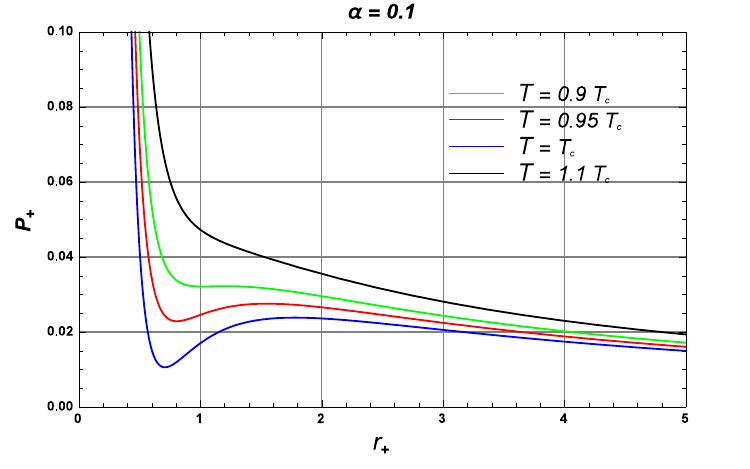}
\includegraphics[width=.525\linewidth]{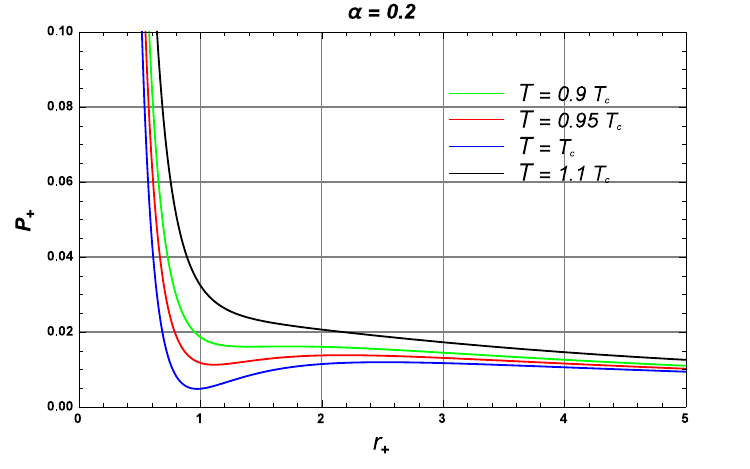}
\end{tabular}
\caption{The plot of pressure Vs. $r_+$ with different value of temperature $T$ for $\alpha=0.1$ and $\alpha=0.2$ with fixed value of $e$. }
\label{fig:1}
\end{figure*}

It is a matter of calculation to obtain the critical radius, critical temperature and critical  pressure at the inflection points by using the following relations:
\begin{equation}
\frac{\partial P_+}{\partial r_+}=0,\qquad \qquad \frac{\partial ^2P_+}{\partial r_+^2}=0.
\label{pv5}
\end{equation}
This relation ({\ref{pv5}}) together with the EoS  simplifies to   the following expression:
\begin{eqnarray}
r_+^8(r_+^2-12\alpha)-2e^6(5r_+^4+54r_+^2\alpha+16\alpha^2)-6e^3(3r_+^7+40r_+^5\alpha +112r_+^3\alpha^2)=0.
\label{pv6}
\end{eqnarray}
One can not solve the above Eq. (\ref{pv6})  analytically. Rather, it is possible to solve it numerically. Here, we calculate the numerical values of   critical radius $r_C$, critical pressure $P_C$ and temperature  $T_C$. { The curved portion of the isotherm that is cut off by this straight-line correctly indicates what the allowed states would be if the fluid were homogeneous. However, these homogeneous states are unstable, because there is always another mixed state at the same pressure that possesses a lower Gibbs free energy.}

The respective numerical  values  are presented in  Tables \ref{tr10} and   \ref{tr12} corresponding to the different values of $\alpha$ and $e$.
\begin{table}[ht]
 \begin{center}
 \begin{tabular}{ |l | l   | l   | l   | l |   }
            \hline
  \multicolumn{1}{|c|}{ $e$} &\multicolumn{1}{c}{$r_C$}  &\multicolumn{1}{|c|}{$T_C$}  &\multicolumn{1}{c|}{$P_C$} &\multicolumn{1}{c|}{${P_C\,r_C}/{T_C}$}\\
            \hline
            
            \,\,\,\,\,0.1~~ &~~1.112~~ & ~~0.144~~ & ~~0.0327~~ &  ~~0.252~~      \\
            \,\,\,\,\,0.2~~ &~~1.204~~  & ~~0.140~~ & ~~0.0302~~ &  ~~0.259~~    \\
            \,\,\,\,\,0.3~~ &~~1.362~~ & ~~0.132~~ & ~~0.0260~~ &    ~~0.268~~  \\
            \,\,\,\,\,0.4~~ &~~1.552~~  & ~~0.122~~ & ~~0.217~~ &    ~~0.276~~   \\
            \,\,\,\,\,0.5~~ &~~1.760~~  & ~~0.112~~ & ~~0.0179~~ &    ~~0.281~~    \\
         \hline
        \end{tabular}
        \caption{The numerical values of critical radius $r_C$, critical temperature $T_C$, critical pressure $P_C$ and $P_C\,r_C/T_C$ corresponding to various $e$ and fixed $\alpha=0.1$.}
\label{tr10}
    \end{center}
\end{table}
\begin{table}[ht]
 \begin{center}
 \begin{tabular}{| l | l   | l   | l   |  l |  }
  \hline
  \multicolumn{1}{|c|}{ $\alpha$} &\multicolumn{1}{c|}{$r_C$}  &\multicolumn{1}{c|}{$T_C$}  &\multicolumn{1}{c|}{$P_C$} &\multicolumn{1}{|c|}{$P_C\,r_C/T_C$}\\
            \hline
            \,\,\,\,\,0.1 ~~  &~~1.112~~  & ~~0.144~~ & ~~0.0327~~ & ~~0.252~~ \\            
            \,\,\,\,\,0.2~~ &~~1.557~~ & ~~0.102~~ & ~~0.0164~~ &  ~~0.250~~      \\
            \,\,\,\,\,0.3~~ &~~1.903~~  & ~~0.0838~~ & ~~0.0110~~ &  ~~0.249~~    \\
            \,\,\,\,\,0.4~~ &~~2.195~~ & ~~0.0725~~ & ~~0.0082~~ &    ~~0.248~  \\
            \,\,\,\,\,0.5~~ &~~2.453~~  & ~~0.0649~~ & ~~0.0066~~ &    ~~0.247~~   \\
         \hline
        \end{tabular}
        \caption{The numerical values of critical radius $r_C$, critical temperature $T_C$, critical pressure $P_C$ and $P_C\,r_C/T_C$ corresponding to various $\alpha$ and fixed $e$.}
\label{tr12}
    \end{center}
\end{table}

It can be seen that the  critical  radius $r_C$  increases both  with $e$ and $\alpha$. However, the critical temperature $T_C$ and critical pressure $P_C$ decrease with both $e$ and $\alpha$.  Interestingly, in contrast to 
$\alpha$ case, the  universal ratio $P_Cr_C/T_C$ increases with $e$.

\begin{figure*}[ht]
\begin{tabular}{c c c c}
\includegraphics[width=.45\linewidth]{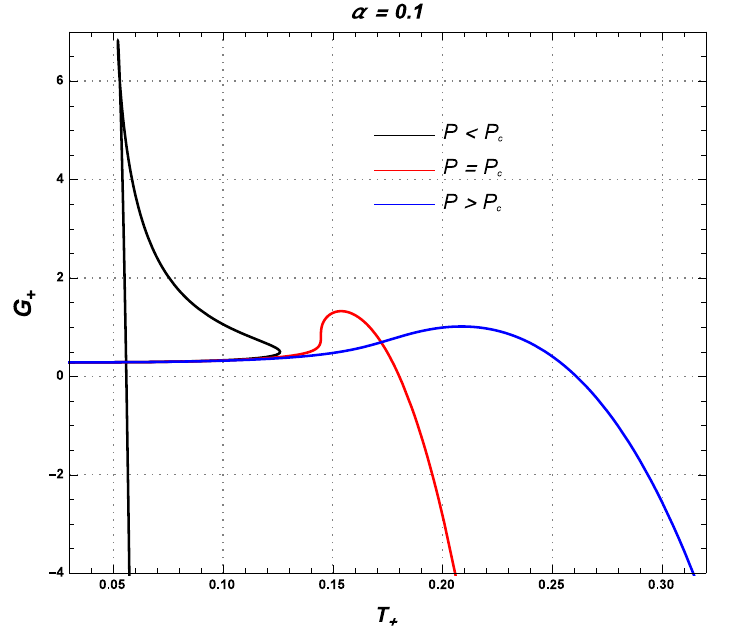}
\includegraphics[width=.45\linewidth]{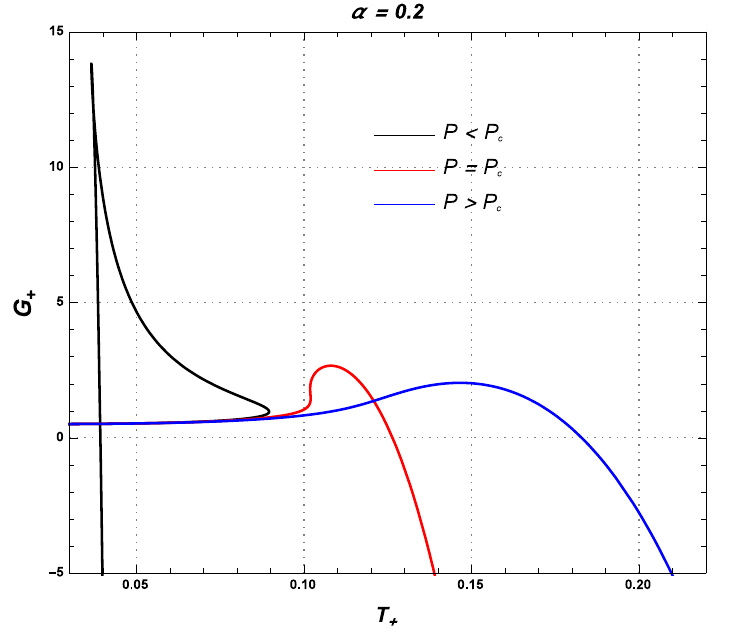}
\end{tabular}
\caption{The plot of  Gibbs free energy $G_+$ versus $T_+$ with different value of GB coupling with fixed value of $e$. }
\label{fig}
\end{figure*}

\begin{figure*}[ht]
\begin{tabular}{c c c c}
\includegraphics[width=.45\linewidth]{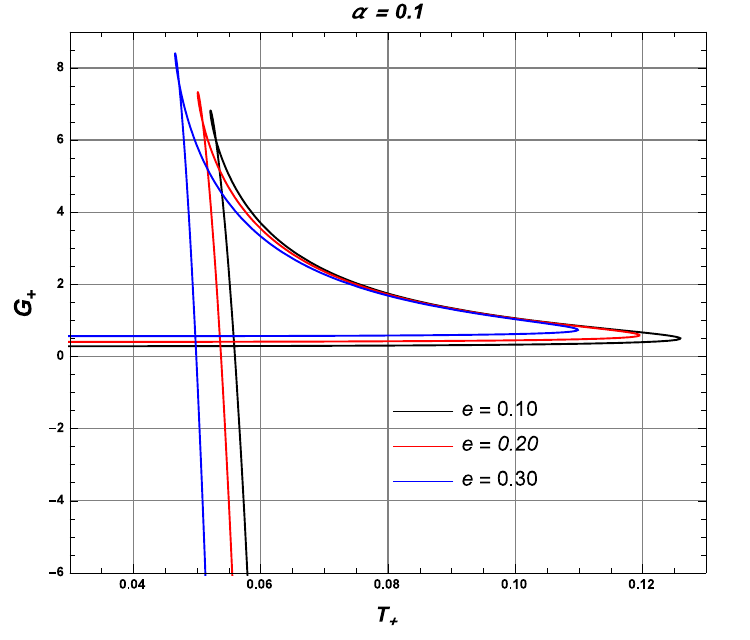}
\includegraphics[width=.45\linewidth]{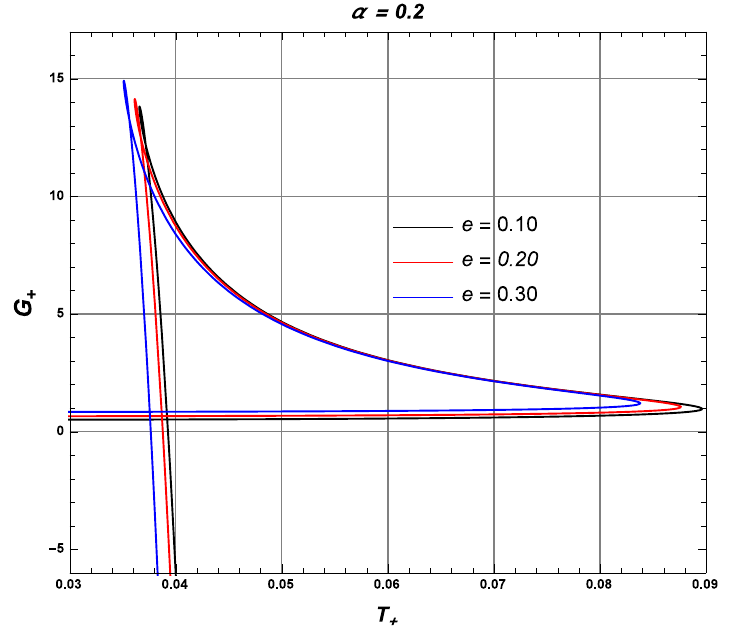}
\end{tabular}
\caption{The plot of  Gibbs free energy $G_+$ Vs. $T_+$ with different value of $e$with fixed value of  GB coupling. }
\label{fig1}
\end{figure*}

\begin{figure*}[ht]
\begin{tabular}{c c c c}
\includegraphics[width=.45\linewidth]{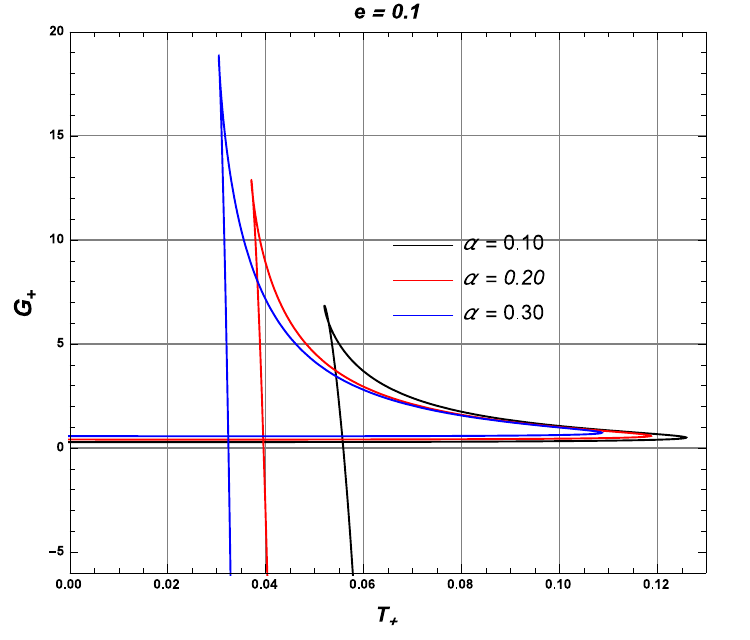}
\includegraphics [width=.45\linewidth]{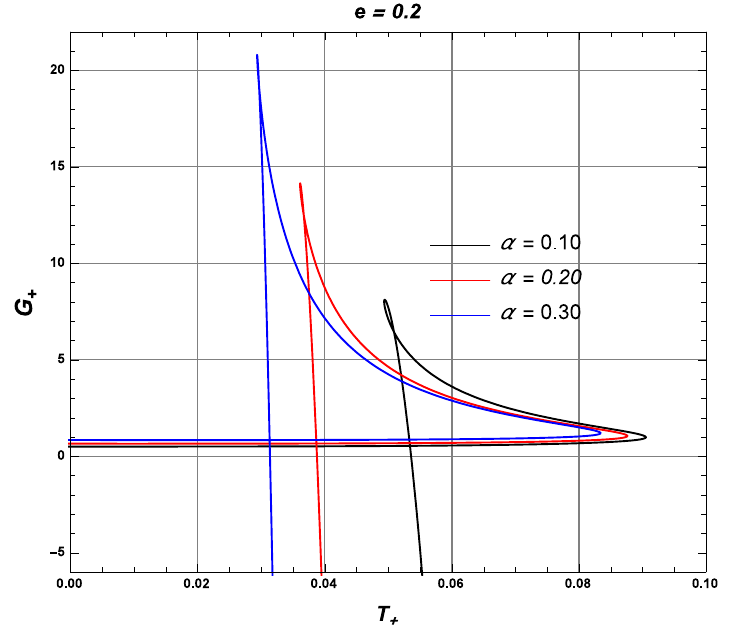}
\end{tabular}
\caption{The plot of Gibbs free energy $G_+$ Vs. $T_+$ with different values of GB coupling  for fixed value $e$. }
\label{fig2}
\end{figure*}

To  study the phase transition and effects  of  $e$ and $\alpha$ on the phase transition of this particular black hole,
  we  plot the Gibbs free energy versus temperature as shown in figures \ref{fig}, \ref{fig1} and \ref{fig2}. 
  From  $G_+-T_+$  plot \ref{fig} with fixed $e$ and $\alpha$  and a certain range of temperature, we observe that there are three kinds of  black holes, namely, small ($P<P_c$), intermediate $(P=P_c)$ and large black hole $(P>P_c)$. Here, the small and large black holes  are more stable than the intermediate ones, since the heat capacity is negative (see  Fig. \ref{fig4}).  We can see that there exists a  transition temperature $T_{\star}$ at which a black hole  transits from one phase to another phase  due to the same free energy.    The value of transit temperatures are $T_{\star}=0.0564$ and $T_{\star}=0.0358$ for $\alpha=0.1$ and $0.2$, respectively. In    Fig. \ref{fig}  isotherms represent the first order phase transition at $T_+<T_{\star}$ and second order  phase transition at $T_+=T_{\star}$, which is obtained from the free energy diagram (see Fig. \ref{fig}. At $ T_+<T_{\star} $ the small black hole occurs and at $ T_+>T_{\star} $  large black hole occurs due to small free energy.

  In $G_+-T_+$  plots, the appearance of characteristic swallow tail shape   shows the phase transition point.  In the left panel of Fig. \ref{fig}, we see that swallow tail shape occurs at $P<P_c$  for the first order phase transition and $P=P_c=0.0327$ for $\alpha=0.1$ and $P=P_c=0.0164$ for $\alpha=0.2$ for the second order phase transition.

 Figs. \ref{fig1} and  \ref{fig2} show    the effects of  $e$ and $\alpha$ on isobars.  The isobars increase with both $e$ and $\alpha$   and the gap between the two sub-critical isobars increases as well.  The sub-critical isobar is the regime where the phase transition occurs.  It is  worth    mentioning  that the  Gibbs free energy in two phases is approximately constant and is less affected by $e$ and $\alpha$.   Interestingly, the universal relation $P_c\,r_c/T_c$  increases with  $e$ and decreases with the GB coupling parameter. This reflects that the effects of electric charge $e$ and  GB coupling are opposite to each other. 
 
 { The black hole are thermodynamically stable state when it has lowest Gibbs free energy. The behaviour of Gibbs free energy ($G_+$) is the function of temperature ($T_+$) for ($P < P_c, P=P_c$) and ($P>P_c$). The plot of Gibbs free energy develops swallow tail structure when ($P>P_c$)   which infers the first order phase transition and the swallow tail structure  disappears (first order phase transition terminates) corresponding to the critical pressure $P_c$, which infers the second order phase transition. There is no phase transition, when thermodynamic pressure is larger than the critical pressure $P_c$. The black hole transits from one phase to another phase due to the same free energy and the corresponding temperature is transition temperature.
 
 }
  \section{Phase Structure in the framework of AdS/CFT Correspondence} 
In order to study the phase structure from the holography point of view, we first  write the  Hawking temperature of a given  black hole in terms of entropy   as follows, 
\begin{equation}
T_+=\frac{{8\pi} }{ (3S_+)^{\frac{1}{3}}( {4\pi e^3+3S_+})[(3S_+)^{\frac{2}{3}}+4 (4\pi)^{\frac{2}{3}}\alpha]}\left[\frac{1}{l^2}\left(\frac{3S_+}{4\pi}\right)^{\frac{7}{3}}+  \left(\frac{3S_+}{4\pi}\right)^{\frac{5}{3}}- e^3\left(\left(\frac{3S_+}{4\pi}\right)^{\frac{2}{3}}+4\alpha\right)  \right].
\end{equation} 

\begin{figure*}[ht]
\begin{tabular}{c c c c}
\includegraphics[width=.35\linewidth]{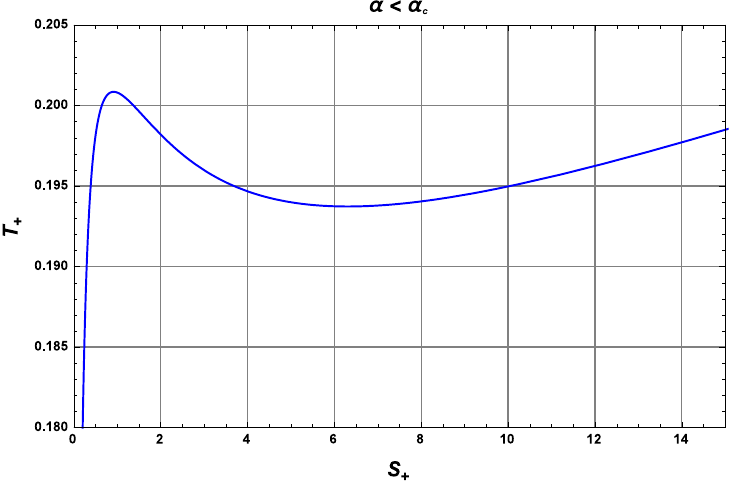}
\includegraphics[width=.35\linewidth]{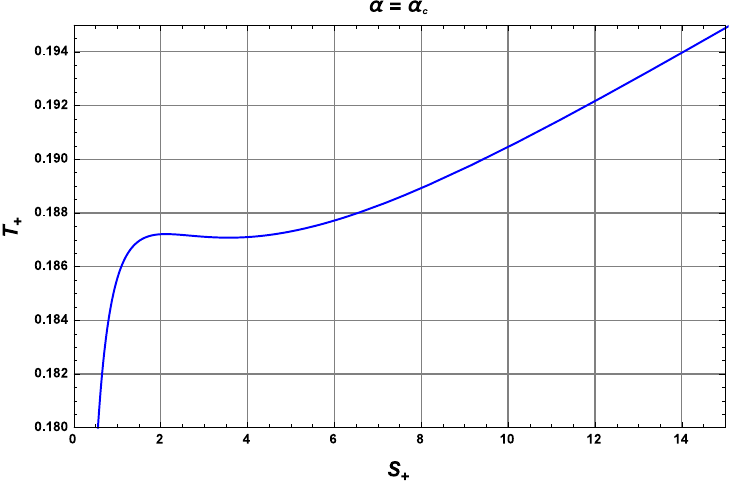}
\includegraphics[width=.35\linewidth]{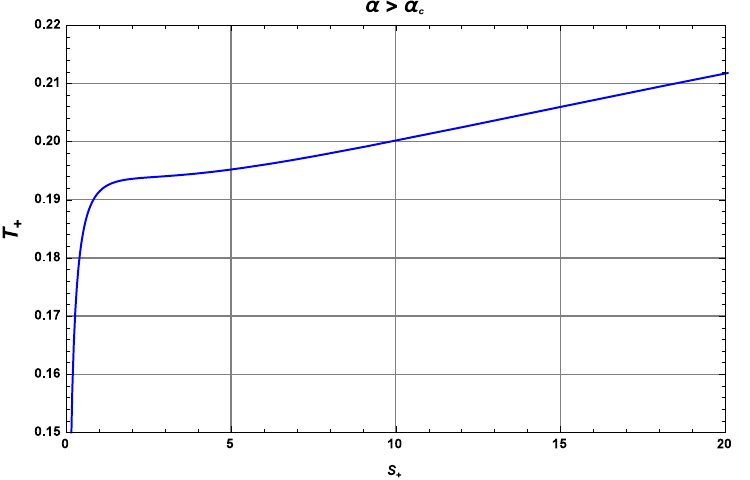}
\end{tabular}
\caption{The plot of    $T_+$ Vs. $S_+$. Here we set the parameter $e=0.1,l=1$ and
the critical value of GB parameter ($\alpha_c$) = 0.0495. }
\label{fig10}
\end{figure*}

Furthermore,  to obtain the critical radius, critical temperature and critical
the temperature at the inflection points, one can utilize   the following relations:
\begin{eqnarray}
\frac{\partial T_+}{\partial S_+} =0, \ \ \ \ \frac{\partial^2 T_+}{\partial S^2_+} =0.
\end{eqnarray}
 In Fig. \ref{fig10}, we plot the scalar isocharges in the $T _+-  S_+$ plane for   $\alpha_c$ = 0.0495.  
According to  AdS/CFT correspondence, Ryu and Takayanagi  \cite{rt,rt1} presented an elegant way to calculate the holographic entanglement entropy which is given by the following relation:
\begin{equation}
S_+=\frac{\mbox{Area of horizon}}{4G}.
\end{equation} 
The holographic entanglement entropy has the following form \cite{li}: 
\begin{eqnarray}
S_+=\pi \int^{\phi_0}_0 r^2\sin^2 \phi  \sqrt{r^2+\frac{1}{f(r)}\left(\frac{dr}{d\phi}\right)^2}d\phi.\label{222}
\end{eqnarray}
Here, we considered $\phi=\phi_0$ as
entangling surface and choose the values of $\phi_0: 0.35, 0.42$ and
$0.50$. 

\begin{figure*}[ht]
\begin{tabular}{c c c c}
\includegraphics[width=.35\linewidth]{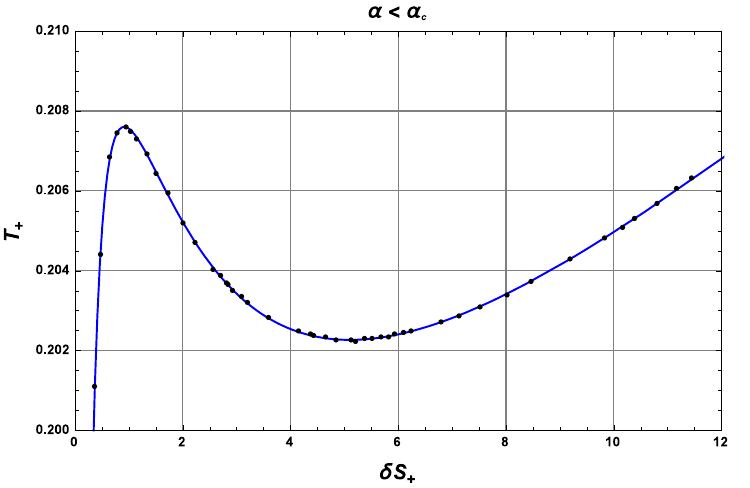}
\includegraphics[width=.35\linewidth]{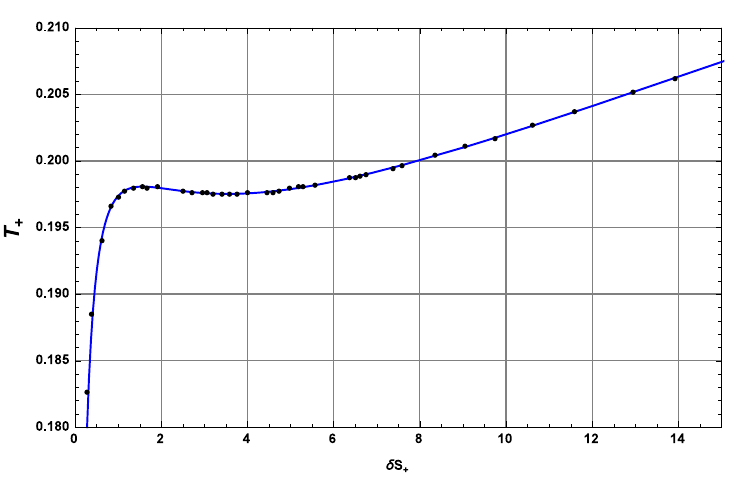}
\includegraphics[width=.35\linewidth]{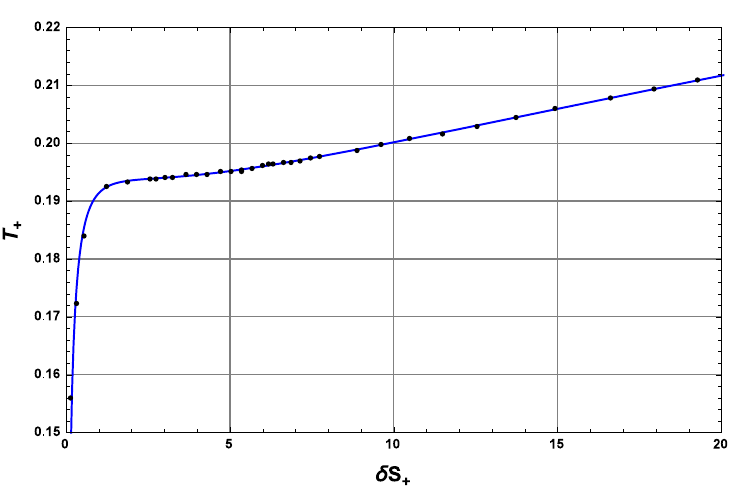}
\end{tabular}
\caption{The plot of    $T_+$ Vs. $S_+$. Here we set the parameter $e=0.1,l=1,\phi_0=0.35$ and $\phi_c=0.0495$. }
\label{fig15}
\end{figure*}

We get the numeric result of $r(\phi)$ with a boundary conditions $r'(0)=0$ and $r(0)=r_0$.
 In order to regularize entanglement entropy, we again integrate $S$ in Eq. (\ref{222}) up to cut-off (which is close to $\phi_0$), and subtract the pure AdS entanglement entropy (denoted by $S_+'$) with a same entangling surface $\phi_0$ at the boundary. Here, the corresponding regularized entanglement entropy is denoted by  $S_+ = S_+-S_+'$.  In Fig. \ref{fig15}, we plot the $T _+- \delta S_+$ plane for  $\phi_0 = 0.35$
 as depicted by the  dotted curve. 
 
With the help of comparative analysis as given in Fig. \ref{fig15}, we find that the entanglement entropy also presents a Van der Waals-like phase transition.
 
\section{Critical Exponents}  \label{sec6}

In this section, we discuss the critical exponents of the fluid behavior of 
black holes. Since, the  Van der Waals-like phase transition is described  by the  critical exponents $\alpha$, $\beta$, $\gamma$ and $\delta$ near the critical point  which shed light on  the nature of the  heat capacity $C_+$,   order parameter $\eta$,  the isothermal compressibility $\kappa_T$ and   the critical isotherm $|P_+-P_C|_{T_C}$, respectively. 
These quantities  are defined by
\begin{eqnarray}
&&C_+=T_+\frac{\partial S_+}{\partial T_+}\propto|t|^{-\alpha},\qquad\qquad \eta=V_l-V_s \propto |t|^{\beta},\\
&&\kappa_T=-\frac{1}{V}\frac{\partial V}{\partial P_+}\propto|t|^{-\gamma},\quad\qquad |P_+-P_C|_{T_C}\propto |V-V_C|^{\delta},
\end{eqnarray}
where $t=\frac{T_+-T_C}{T_C}$,  $V_l$  and  $V_s$  are  the volumes of a large-sized and the   small sized
black holes, respectively. Now, we use the following definitions:
\begin{equation}
p=\frac{P_+}{P_c},\qquad\qquad \tau=\frac{T_+}{T_c}\qquad\text{and}\qquad \nu =\frac{V}{V_c},
\end{equation}  
Substituting these values in Eq. (\ref{pv4}), we get the   EoS in terms of dimensionless parameters as
\begin{equation}
p=\frac{1}{\rho_C} \frac{\tau}{\nu}+8\alpha\frac{1}{\rho_CV_C^2} \frac{\tau}{\nu}-\frac{1}{P_CV_C^2}\frac{1}{2\pi \nu^2}+\frac{1}{V_C^4 P_C}\frac{\alpha+\nu^2}{\pi \nu^4}.
\end{equation}
Let us assume that the law of the critical exponents defined as $p=\frac{1}{\rho_C}\frac{\tau}{\nu}+h(\nu)$, where $\rho_c$ is the critical ratio and $h(\nu)$ is the correction  term. Now, expanding this equation near the following critical points:
\begin{equation}
\tau=t+1,\qquad\qquad \nu=(1+\omega)^{1/z},
\label{es1}
\end{equation}
and substituting these values of $\nu$ and $\omega$ in Eq (\ref{es1}), we have 
\begin{equation}
\frac{1}{\rho_C}+h(1)=1,\qquad\rho_ch'(1)=1,\quad\text{and}\quad\rho_ch''(1)=-2.
\end{equation}
In order to obtain the critical exponents, we expand  the EoS near the critical points in the following manner:
\begin{equation}
p=1+A_{10} t+A_{11}t \omega+A_{03}\omega^3+\mathcal{O}(t\omega^2,\omega^4),
\end{equation}
comparing  Eq. (40) and Eq. (37), the coefficients of critical exponents are
\begin{eqnarray}
&&A_{10}=\frac{1}{\rho_C},\qquad\qquad A_{11}=\frac{1}{z\rho_C} \quad\text{and}\quad A_{03}= \frac{1}{z^3}\left(\frac{}{}-\frac{h^3(1)}{6}\right).
\end{eqnarray}
Now, we note that for a phase transition  from small to large black holes
  the pressure and temperature remain constant while volume changes. In this case, the EoS holds and this gives 
 \begin{eqnarray}
p= 1+A_{10} t+a_{11}t \omega_s+A_{03}\omega_s^3 
= 1+A_{10} t+A_{11}t \omega_l+A_{03}\omega_l^3.
\end{eqnarray} 
Upon simplification, we have
\begin{equation}
A_{11}t (\omega_l^2-\omega_s^2)+A_{03}(\omega_l^3-\omega_s^3)=0.
\end{equation}
Moreover, it also follows Maxwell's area law during the phase transition, i.e.,
\begin{equation}
\int_{\omega_s}^{\omega_l}\omega  dp=A_{11}t (\omega_l^2-\omega_s^2)+\frac{3}{2}A_{03}(\omega_l^3-\omega_s^3)=0.
\end{equation}
Solving for the above equations, we obtain a non-trivial solution
\begin{equation}
\omega_l=-\omega_s=\sqrt{\frac{-A_{11}t}{A_{03}}}.
\end{equation}
With the obtained  solution,  the order parameter, isothermal compressibility and the shape of  critical isotherms read, respectively,
\begin{eqnarray}
&&\eta=V_C(\omega_l-\omega_s)=2V_C \omega_l\propto\sqrt{-t},\\
&&\kappa_T=-\left. \frac{1}{V}\frac{\partial V}{\partial P_+}\right|_T\propto\frac{1}{P_C}\frac{1}{t},\\
&& p-1 \propto - \omega^3, \qquad\qquad\qquad\qquad  \text{at}\  t=0.
\end{eqnarray}
Consequently, the four critical exponents are $\alpha=0,\beta=\frac{1}{2},\gamma=1,\delta=3$, which exactly  match with the mean field theory.
\section{Final remarks and perspectives}\label{sec7}
We have found a new exact  Bardeen AdS black hole in   EGB gravity coupled with nonlinear electrodynamics. Our solution interpolates to the Boulware-Deser  black hole \cite{7} when $e=0$, 
$5D$ Bardeen black hole when $\alpha\to 0$ \cite{Ali:2018boy} and 
the schwarzschild-Tangherilin black hole in the limit of $e=0,\alpha \to 0$.  
We have found that the obtained black hole has a central de Sitter core. 
The horizon structure is discussed numerically as it is not possible to 
solve it analytically. The black hole has two horizons. The curvature  scalars are not singular at the centre which justifies that the obtained solution is regular.

 Furthermore, we have studied the thermodynamics of the obtained black holes by computing temperature,  entropy,  heat capacity and free energy.
 We have found that this black hole satisfies the modified first law of thermodynamics. The local and global stability are also emphasized by
 studying the diagram of heat capacity and Gibbs free energy. The phase transition and critical points are also found for this black hole solution.
 Finally, we found that the   critical exponents calculated here are in full agreement  with the mean field theory.  

\begin{acknowledgements}  
This research has been/was/is funded by the Science Committee of the Ministry of Science and Higher Education of the Republic of Kazakhstan (Grant No. AP09058240).
 D.V.S.  thanks University Grant Commission for the Start Up Grant No.30-600/2021(BSR)/1630. 
\end{acknowledgements}

\section*{Data Availability Statement}
No Data is associated with the manuscript.
\section*{Conflict of Interest Statement}
The authors declare that they have no known competing financial interests or personal relationships that could have
appeared to influence the work reported in this paper.


\begin{thebibliography}{99}

\bibitem{lav} D. Lovelock,  J. Math. Phys. {\bf 12} (1971) 498.
\bibitem{lav1} D. Lovelock,  J. Math. Phys. {\bf 13} (1972) 874.
\bibitem{lav2} N. Deruelle and L. Farina-Busto, Phys. Rev. D {\bf 41} (1990) 3696.
\bibitem{noj} S. Nojiri, S. D. Odintsov and V. K. Oikonomou, Phys. Rept. \textbf{692} (2017) 1.




\bibitem{lan} C. Lanczos, Ann. Math. {\bf 39}, 842 (1938).
\bibitem{zwi}B. Zwiebach, Phys. Lett. B {\bf 156} (1985) 315.
\bibitem{ds}S. Upadhyay and D. V. Singh, Eur. Phys. J. Plus {\bf 137} (2022)  383.
\bibitem{ds1}D. V. Singh, B. K. Singh and S. Upadhyay, Annals Phys.  {\bf 434} (2021) 168642.




 

\bibitem{Sakharov:1966}
A.D.~Sakharov, Sov.\ Phys.\ JETP  {\bf 22}, 241 (1966).
	
\bibitem{Gliner:1966}
E.B.~Gliner, Sov.\ Phys.\ JETP  {\bf 22}, 378 (1966).
\bibitem{Bardeen:1968}
J.~Bardeen,
in {\it Proceedings of GR5} (Tiflis, U.S.S.R., 1968).
\bibitem{AGB}   E.~Ayon-Beato and A.~Garcia,
Gen.  Rel.  Grav.  {\bf 31}, 629 (1999).
\bibitem{AGB1}
 E. Ayon-Beato and A. Garcia, Gen. Rel. Grav. {\bf 37}, 635
(2005).
\bibitem{ABG99}E. Ayon-Beato, A. Garcia, Phys. Lett. B {\bf 493}, 149 (2000).
\bibitem{AyonBeato:1998ub}
E.~Ayon-Beato and A.~Garcia, Phys.\ Rev.\ Lett.\  {\bf 80}, 5056 (1998).

\bibitem{Ansoldi:2008jw} S.~Ansoldi, arXiv:0802.0330 [gr-qc].

	\bibitem{Bronnikov:2000vy} K. A.~Bronnikov,  	Phys.\ Rev.\ D {\bf 63}, 044005 (2001).
	 
	\bibitem{Zaslavskii:2009kp} 	O. B.~Zaslavskii, Phys.\ Rev.\ D {\bf 80}, 064034 (2009).
 
	\bibitem{Lemos:2011dq}
	J. P. S.~Lemos and V. T.~Zanchin,.
	Phys.\ Rev.\ D {\bf 83}, 124005 (2011)	.
\bibitem{hc} H.~Culetu, arXiv:1408.3334v1 [gr-qc].
\bibitem{lbev}  L.~Balart and E.~C.~Vagenas,  Phys.  Lett.  B {\bf 730}, 14 (2014).
\bibitem{Balart:2014cga} L.~Balart and E.~C.~Vagenas, Phys.\ Rev.\ D {\bf 90}, 124045 (2014).
\bibitem{Xiang} L.~Xiang, Y.~Ling and Y.~G.~Shen, Int.\ J.\ Mod.\ Phys.\ D {\bf 22}, 1342016 (2013).

\bibitem{singh}
  D.~V.~Singh and N.~K.~Singh,
  Annals Phys.  {\bf 383}, 600 (2017).
\bibitem{dvs}
  D.~V.~Singh, M.~S.~Ali and S.~G.~Ghosh, 
 Int. J. Mod. Phys. D \textbf{27} (2018)   1850108.
\bibitem{fr1}
S. Fernando,   Int. J. Mod. Phys. D {\bf 26}, 1750071 (2017).
 
\bibitem{bks1}
D.~V.~Singh, S.~G.~Ghosh and S.~D.~Maharaj, 
Nucl. Phys. B \textbf{981} (2022)  11585.

\bibitem{bks2}
D.~V.~Singh, V.~K.~Bhardwaj and S.~Upadhyay, 
Eur. Phys. J. Plus \textbf{137} (2022)   969.
 \bibitem{bks3}

D.~V.~Singh, A.~Shukla and S.~Upadhyay, 
Annals Phys. \textbf{447} (2022), 169157.
\bibitem{Singh:2022xgi}
D.~V.~Singh, S.~G.~Ghosh and S.~D.~Maharaj, 
Nucl. Phys. B \textbf{981} (2022)  115854.
\bibitem{Singh:2022dth}
D.~V.~Singh, V.~K.~Bhardwaj and S.~Upadhyay, Eur. Phys. J. Plus \textbf{137} (2022)  969.
\bibitem{Hendi:2015soe}
S.~H.~Hendi, S.~Panahiyan and B.~Eslam Panah, Prog. Theor. Exp. Phys. \textbf{2015} (2015)  103E01.
\bibitem{Hendi:2015xya}
S.~H.~Hendi, A.~Sheykhi, S.~Panahiyan and B.~Eslam Panah, 
Phys. Rev. D \textbf{92} (2015)   064028.



\bibitem{Kumar:2020bqf}
A.~Kumar, D.~V.~Singh and S.~G.~Ghosh, 
Annals Phys. \textbf{419}, 168214 (2020).


\bibitem{Singh:2019wpu}
D.~V.~Singh, S.~G.~Ghosh and S.~D.~Maharaj, 
Annals Phys. \textbf{412}, 168025 (2020).

 
\bibitem{Kumar:2018vsm}
A.~Kumar, D.~V Singh and S.~G.~Ghosh, 
Eur. Phys. J. C \textbf{79}, 275 (2019).

 
\bibitem{Ghosh:2018bxg}
S.~G.~Ghosh, D.~V.~Singh and S.~D.~Maharaj, 
Phys. Rev. D \textbf{97},  104050 (2018).

\bibitem{Ghosh:2020tgy}
S.~G.~Ghosh, A.~Kumar and D.~V.~Singh, 
Phys. Dark Univ. \textbf{30}, 100660 (2020).

\bibitem{Singh:2020mty}
D.~V.~Singh, R.~Kumar, S.~G.~Ghosh and S.~D.~Maharaj,
Annals Phys. \textbf{424} (2021) 168347.

\bibitem{Singh:2020nwo}
D.~V.~Singh, S.~G.~Ghosh and S.~D.~Maharaj, 
Phys. Dark Univ. \textbf{30} (2020), 100730.
\bibitem{Singh:2020xju}
D.~V.~Singh and S.~Siwach, 
Phys. Lett. B. {\bf 408} 135658 (2020).

\bibitem{Singh:2020rnm}
B.~K.~Singh, R.~P.~Singh and D.~V.~Singh, 
Eur. Phys. J. Plus \textbf{135} (2020)  862.
\bibitem{Bambi}  C.~Bambi and L.~Modesto,  Phys.\ Lett.\ B {\bf 721}, 329 (2013).
\bibitem{Ghosh:2014pba} 
S.~G.~Ghosh,
Eur.\ Phys.\ J.\ C {\bf 75}, 532 (2015).
\bibitem{Toshmatov:2014nya} B.~Toshmatov, B.~Ahmedov, A.~Abdujabbarov and Z.~Stuchlik,
Phys.\ Rev.\ D {\bf 89}, 104017 (2014).
\bibitem{Ghosh:2014hea} S.~G.~Ghosh and S.~D.~Maharaj, Eur.  Phys.  J.  C {\bf 75},   7 (2015).
\bibitem{Neves:2014aba}
J.~C.~S.~Neves and A.~Saa,
Phys.\ Lett.\ B {\bf 734}, 44 (2014).

 \bibitem{noj1}S. D. Odintsov, V. K. Oikonomou and F. P. Fronimos,
 Nucl. Phys. B \textbf{958} (2020) 115135.
 \bibitem{noj2}V. K. Oikonomou, Class. Quant. Grav. \textbf{38} (2021)   195025.
 \bibitem{noj3} G. G. L. Nashed, S. D. Odintsov and V. K. Oikonomou,
 Symmetry \textbf{14} (2022)   545.
 \bibitem{noj4}V. K. Oikonomou, Astropart. Phys. \textbf{141} (2022) 102718.
 
\bibitem{sud1}S. H. Hendi, B. E. Panah andS. Panahiyan, J. High Energy Phys. {\bf 11} (2015) 157.
\bibitem{sud2}S. Upadhyay, Gen. Rel. Grav. {\bf 50} (2018) 128.
\bibitem{sud3} S. Upadhyay, S. H. Hendi, S. Panahiyan and B. E. Panah, Prog. Theor. Exp. Phys. \textbf{2018}, { 093E} (2018).
\bibitem{sud4}B. Pourhassan, S. Upadhyay, H. Saadat and H. Farahani, Nucl. Phys. B {\bf 928} (2018) 415.
\bibitem{sud5} S. Upadhyay, Phys. Lett. B {\bf 775} (2018) 130.
\bibitem{Kubiznak:2016qmn}
D.~Kubiznak, R.~B.~Mann and M.~Teo, Class. Quant. Grav. \textbf{34} (2017) 063001.



\bibitem{h1}
S. Hawking and D. Page, Commun. Math. Phys. \textbf{87}, 577 (1983).
\bibitem{w1}
E. Witten, Adv. Theor. Math. Phys. \textbf{2}, 505 (1998).
\bibitem{c1}
A. Chamblin, R. Emparan, C. V. Johnson and R. C. Myers, Phys. Rev. D \textbf{60} (1999) 104026.
\bibitem{c2}
A. Chamblin, R. Emparan, C. V. Johnson and R. C. Myers, Phys. Rev. D \textbf{60} (1999) 06401.

\bibitem{Hansen:2016ayo}
D.~Hansen, D.~Kubiznak and R.~B.~Mann,
JHEP \textbf{01} (2017), 047.

\bibitem{Hendi:2018sbe}
S.~H.~Hendi and M.~Momennia, 
Phys. Lett. B \textbf{777} (2018) 222.

\bibitem{Hendi:2014kha}
S.~H.~Hendi, S.~Panahiyan and B.~Eslam Panah,
Int. J. Mod. Phys. D \textbf{25} (2015)   1650010
\bibitem{Hennigar:2016ekz}
R.~A.~Hennigar, E.~Tjoa and R.~B.~Mann, 
JHEP \textbf{02} (2017)  070.
\bibitem{rps}
R. P. Singh, B. K. Singh, B. R. K. Gupta and S. Sachan, Can. J. Phys. \textbf{100}, 1 (2022).

\bibitem{Ali:2023ppg}
M.~S.~Ali, S.~G.~Ghosh and A.~Wang,
Phys. Rev. D \textbf{108} (2023) no.4, 044045.

\bibitem{Kumar:2023gjt}
A.~Kumar, A.~Sood, J.~K.~Singh, A.~Beesham and S.~G.~Ghosh,
Phys. Dark Univ. \textbf{40} (2023), 101220.


\bibitem{Sood:2022fio}
A.~Sood, A.~Kumar, J.~K.~Singh and S.~G.~Ghosh,
Eur. Phys. J. C \textbf{82} (2022) no.3, 227.

\bibitem{Abdusattar:2023xxs}
H.~Abdusattar,
Phys. Dark Univ. \textbf{40} (2023), 101228.
\bibitem{Abdusattar:2023fdm}
H.~Abdusattar,
Eur. Phys. J. C \textbf{83} (2023) no.7, 614.
\bibitem{Abdusattar:2023pck}
H.~Abdusattar,
JHEP \textbf{09} (2023), 147.

\bibitem{Abdusattar:2023hlj}
H.~Abdusattar, S.~B.~Kong, H.~Zhang and Y.~P.~Hu,
Phys. Dark Univ. \textbf{42} (2023), 101330.

\bibitem{sd1}
D.~V.~Singh, S.~Upadhyay and M.~S.~Ali, 
Int. J. Mod. Phys. A \textbf{37} (2022)   2250049.
\bibitem{a1}
H. Salazar, A. Garcia and J. Plebanski, J. Math. Phys {\bf 28} 2171 (1987).
\bibitem{gra}M. Born and L. Infeld,   Proc. R. Soc. Lond. A \textbf{144}, 425 (1934).
\bibitem{shu}A. Kumar and S. G. Ghosh, Nucl. Phys. B \textbf{987} (2023) 116089.
 \bibitem{shu1}S. G. Ghosh, D. V. Singh, R. Kumar and S. D. Maharaj, Annals Phys. \textbf{424} (2021) 168347.
 
 
 
 
  \bibitem{7}
D. G. Boulware and S. Deser, Phys. Rev. Lett. {\bf 55}, 2656 (1985).
\bibitem{Ali:2018boy}
M.~S.~Ali and S.~G.~Ghosh, 
Phys. Rev. D \textbf{98} (2018) 084025.

\bibitem{ma14}
M. Ma and R.  Zhao, Class. Quant. Grav. {\bf{31}},  245014 (2014).

\bibitem{M2} 
  R.~V.~Maluf and J.~C.~S.~Neves, 
  Phys. Rev.  D {\bf 97}, 104015 (2018).
  
\bibitem{wald93}
R. M.  Wald, Phys. Rev. D {\bf 43}, R3427 (1993).

\bibitem{rt} S. Ryu and T. Takayanagi, Phys. Rev. Lett. \textbf{96} (2006) 181602.
\bibitem{rt1} S. Ryu and T. Takayanagi, J. High Energy Phys. \textbf{0608} (2006) 045. 

\bibitem{li} H.-L. Li, S.-Z. Yang and X.-T. Zu, Phys. Lett. B \textbf{764}  (2017) 310. 



\end{thebibliography}
\end{document}